# Multipartite entangled state representation and its squeezing transformation

Sergio Albeverio [a] [1], Shao-Ming Fei [a,b] [2], Tong-qiang Song [c] [3]

[a] Institut fur Angewandte Mathematik, Universitat Bonn, D-53115
[b] Department of Mathematics, Capital Normal University, Beijing 100037
[c] Department of Physics, Ningbo University, Ningbo 315211

**Abstract**

The multipartite entangled state $\left|p,\chi_2,\chi_3,\ldots,\chi_n\right\rangle$ for the total momentum and relative coordinates of n particles is constructed. The corresponding quantum mechanical operator with respect to the classical transformation $p \to e^{\lambda_1} p$, $\chi_i \to e^{\lambda_i}\chi_i$, $i = 2,\ldots,n$, in the state $\left|p,\chi_2,\chi_3,\ldots,\chi_n\right\rangle$ is investigated.

## 1.Introduction

In recent years quantum entangled states and entanglement have attracted much attention of both physicists and mathematicians due to their potential applications in quantum communication, quantum teleportation and quantum state engineerings [1-7]. Einstein, Podolsky and Rosen (EPR) first introduced the concept of quantum entanglement by using the commutative property of two particles' relative coordinates and total momentum [8]. To show the incompleteness of quantum mechanics, EPR considered the position wave function $\psi(x_1,x_2) = C\delta(x_1 - x_2 - u)$, which describes perfectly correlated position $(x_1 - x_2 = u)$ and momenta $(p_1 + p_2 = 0)$. In Refs.[9-11], the common eigenstate $|\eta>$ of commutative operators ($\hat{X}_1 - \hat{X}_2, \hat{P}_1 + \hat{P}_2$) in two-mode Fock space is constructed, i.e., $|\eta >= \exp\left[-\frac{1}{2}|\eta|^2 + \eta a_1^+ - \eta^* a_2^+ + a_1^+ a_2^+\right]|00>$, where $\eta = \eta_1 + i\eta_2$ is a complex number, $a_i$ and $a_i^+$, ($i = 1,2$), are the two-mode Bose annihilation and creation

[1] e-mail: albeverio@uni-bonn.de
[2] e-mail:fei@uni-bonn.de
[3] e-mail: stq504@nbip.net.cn

operators , given by $\hat{X}_i = (a_i + a_i^+)/\sqrt{2}$, $\hat{P}_i = (a_i - a_i^+)/\sqrt{2}i$. It has been shown that $|\eta>$ is able to make up a two-particle EPR entangled state representation and is very convenient in treating many problems such as quantum teleportation, quantum dense coding, fractional Fourier transformation etc. (see e.g. the review articles [12-13] ).

It is quite crucial to manipulate multipartite entanglements. In [14] four qubits entanglement has been established. Multipartite entanglement of continuous variables (CV) is a fascinating alternative to the one for discrete variables because of the high efficiency and unconditionalness of quantum optical implementation ( see e.g. the recent review article [15] ). van Loock and Braunstein have shown that one single-mode squeezed state distributed among n parties using linear optics suffices to produce a truly CV n-partite entangled state for any nonzero squeezing and arbitrarily many parties[16]. Recently, Pfister and coworkers have shown theoretically that concurrent interactions in a second-order nonlinear medium placed inside an optical resonator can generate CV multipartite entanglement between the resonator modes[17].

In this paper, we shall construct the common eigenvector $|p,\chi_2,\chi_3,...,\chi_n>$, for simplicity denoted as $|p,\chi>$, of n compatible operators $\hat{Q}_i = \hat{X}_1 - \hat{X}_i$, $(i = 2,3,...,n)$, which are the relative coordinate operators among the n particles, and the total momentum operator $\hat{P} = \sum_{k=1}^{n} \hat{P}_k$, which gives rise to a multipartite entangled state representation. We study the quantum mechanical operator corresponding to a classical transformation p $\rightarrow e^{\lambda_1} p$ , $\chi_2 \rightarrow e^{\lambda_i} \chi_i$ , $i = 2,...,n$ , in the state $\left|p,\chi_2,\chi_3,...,\chi_n\right\rangle$ , Three n-mode realizations of SU(1,1) Lie algebra as well as the corresponding n-mode squeezed states are obtained. We also discuss their quantum optical implementation.

## 2.n-mode entangled state representation

It is easily shown that the n operators $\hat{Q}_i = \hat{X}_1 - \hat{X}_i$, $(i = 2,3,...,n)$, which are the relative coordinate operators among the n particles, and the total momentum operator $\hat{P} = \sum_{k=1}^{n} \hat{P}_k$ are mutually commutative, and thus have common eigenvector. By an explicit calculation we obtain their common eigenvector $|p,\chi_2,\chi_3,...,\chi_n> \equiv |p,\chi>$

$$|p,\chi>= \frac{1}{\sqrt{n}\pi^{n/4}} \exp\left[ -\frac{1}{4} y_j y_j^* + y_j a_j^+ + \frac{2}{n} \sum_{k>l=1}^{n} a_l^+ a_k^+ - \frac{n-2}{2n} a_j^+ a_j^+ \right] |vacuum>, \tag{1}$$

here and hereafter the repeated index means summation from 1 to n, $|vacuum>$ denotes

n-mode vacuum, and

$$y_1 = \frac{\sqrt{2}}{n}(ip + \sum_{k=2}^{n} \chi_n), \quad y_i = y_1 - \sqrt{2}\chi_i, \qquad (i = 2,3,...,n), \tag{2}$$

Especially when $p = \chi_1 = \chi_2 = ... = \chi_n = 0$, Eq.(1) becomes

$$| p,\chi >|_{p=\chi=0} = \frac{1}{\sqrt{n}\pi^{n/4}} \exp\left[\frac{2}{n}\sum_{k>l=1}^{n} a_l^+ a_k^+ - \frac{n-2}{2n} a_j^+ a_j^+\right] | vacuum >$$

$$= \frac{1}{\sqrt{2\pi}} \int_{-\infty}^{\infty} | x >_1 \otimes | x >_2 \otimes | x >_3 \otimes ... \otimes | x >_n dx, \tag{3}$$

where $| x >_i$, $(i = 1,2,3,...,n)$, are coordinate eigenstates

$$| x >_i = \pi^{-1/4} \exp\left(-\frac{1}{2}x^2 + \sqrt{2}xa_i^+ - \frac{1}{2}a_i^{+2}\right) | 0 >_i, \quad \hat{X}_i | x >_i = x | x >_i \tag{4}$$

According to van Loock and Braunstein's method[16], this state can be generated from n squeezed modes of the field emitted by optical parametric oscillators ( OPO's ) below threshold [ i.e. optical parametric amplifiers ( OPA's ) ] and appropriately balanced beam splitters. Let

$$\hat{B}_{ij}(\theta): \begin{cases} a_i \to a_i \cos\theta + a_j \sin\theta \\ a_j \to a_i \sin\theta - a_j \cos\theta \end{cases}, \qquad \theta \in [0,2\pi], \tag{5}$$

$$\hat{N}_{1...n} = \hat{B}_{n-1\ n}(\pi/4)\hat{B}_{n-2\ n-1}(\cos^{-1} 1/\sqrt{3}) \times ... \times \hat{B}_{12}(\cos^{-1} 1/\sqrt{n}). \tag{6}$$

Applying the beam splitter operator $\hat{N}_{1...n}$ to a zero-momentum eigenstate in mode 1 and n-1 zero-position eigenstates in modes 2 through n, we can obtain the n-mode entangled state Eq.(3)

It is easy to prove the eigenvector equations

$$\hat{P} | p,\chi >= p | p,\chi >, \quad \hat{Q}_i | p,\chi >= \chi_i | p,\chi >, \qquad (i = 2,3,...,n) \tag{7}$$

Using the normal ordering form of the vacuum projection operator

$$| 0 >< 0 | =: e^{-a^+ a} :, \tag{8}$$

where : : denotes the normal ordering, and the technique of integration within an ordered product (IWOP) of operators we can prove the completeness relation, i.e.,

$$\int_{-\infty}^{\infty} dp d\chi_2 d\chi_3 ... d\chi_n | p,\chi >< p,\chi |$$

$$= \frac{1}{n\pi^{n/2}} \int_{-\infty}^{\infty} dp d\chi_2 d\chi_3 ... d\chi_n \exp\left[-\frac{1}{2}y_j y_j^* + y_j a_j^+ + \frac{1}{n}\sum_{k>l=1}^{n} a_l^+ a_k^+ - \frac{n-2}{2n} a_j^+ a_j^+\right]$$

$$:\exp\left(-a_j^+ a_j\right):\exp\left[y_j{}^* a_j+\frac{1}{n}\sum_{k>l=1}^{n} a_l a_k-\frac{n-2}{2n}a_j a_j\right]$$

$$=\frac{1}{n\pi^{n/2}}\int_{-\infty}^{\infty} dp d\chi_2 d\chi_3 ... d\chi_n :\exp\left[-\frac{1}{n}(p-\hat{P})^2\right]\exp\left(-\chi' N\tilde{\chi}'\right):=1 \tag{9}$$

where $N$ is a symmetric $(n-1)\times(n-1)$ matrix, $N_{kl}=\begin{cases}(n-1)/n, k=l\\ -1/n, k\neq l\end{cases}$,

$\chi'=(\chi_2-\hat{Q}_2,\chi_3-\hat{Q}_3,...,\chi_n-\hat{Q}_n)$, $\tilde{\chi}'$ denotes the transposition. From Eqs.(7), we can obtain the orthogonality relation

$$<p,\chi\,|\,p'',\chi''>=\delta(p-p'')\delta(\chi_2-\chi_2'')...\delta(\chi_n-\chi_n''). \tag{10}$$

Thus $|\,p,\chi>$ is able to make up a quantum mechanical representation.

Now, we consider the Fourier tansformation of the stae $|\,p,\chi>$ with respect to $\chi_i$, $i=2,...,n$, variables

$$\frac{1}{(2\pi)^{(n-1)/2}}\int_{-\infty}^{\infty}|\,p,\chi>\exp\left[-i(\omega_2\chi_2+\omega_3\chi_3+...+\omega_n\chi_n)\right]d\chi_2 d\chi_3 ... d\chi_n$$

$$=\frac{1}{(2\pi)^{(n-1)/2}\sqrt{n}\pi^{n/4}}\exp\left[-\frac{p^2}{2n}+\frac{i\sqrt{2}p}{n}\sum_{k=1}^{n}a_k^+ +\frac{2}{n}\sum_{k>l=1}^{n}a_l^+ a_k^+ +\frac{2-n}{2n}a_j^+ a_j^+\right]$$

$$\int_{-\infty}^{\infty} d\chi_2 d\chi_3 ... d\chi_n \exp\left[-\chi F\tilde{\chi}+\chi\tilde{v}\right]|\,vacuum>$$

$$=\frac{1}{\pi^{n/4}}\exp\left[-\frac{p^2}{2n}+\frac{i\sqrt{2}p}{n}\sum_{k=1}^{n}a_k^+ +\frac{2}{n}\sum_{k>l=1}^{n}a_l^+ a_k^+ +\frac{2-n}{2n}a_j^+ a_j^+\right]$$

$$\exp\left[\frac{1}{4}vF^{-1}\tilde{v}\right]|\,vacuum>, \tag{11}$$

where $F$ and $F^{-1}$ are symmetric $(n-1)\times(n-1)$ matrices,

$$F_{kl}=\begin{cases}(n-1)/2n, k=l\\ -1/2n, k\neq l\end{cases}, \tag{12}$$

$$(F^{-1})_{kl}=\begin{cases}4, k=l\\ 2, k\neq l\end{cases}, \tag{13}$$

$$\chi=(\chi_2,\chi_3,...,\chi_n), \tag{14}$$

$$\nu = (\nu_2, \nu_3, ..., \nu_n), \qquad \nu_k = \frac{\sqrt{2}}{n}(\sum_{j=1}^{n} a_j^+ - n a_k^+) - i\omega_k, \;\; k = 2,3,...,n, \tag{15}$$

$\tilde{\chi}$ and $\tilde{\nu}$ are the transpositions of $\chi$ and $\nu$, respectively. Eq.(11) further gives rise to

$$\frac{1}{(2\pi)^{(n-1)/2}} \int_{-\infty}^{\infty} | p, \chi > \exp[-i(\omega_2\chi_2 + \omega_3\chi_3 + ... + \omega_n\chi_n)] d\chi_2 d\chi_3 ... d\chi_n$$

$$= \frac{1}{\pi^{n/4}} \exp\left[ -\frac{1}{2n} p^2 - \frac{1}{4}\omega F^{-1}\tilde{\omega} + \frac{i\sqrt{2}p}{n} \sum_{k=1}^{n} a_k^+ - i\sqrt{2}\omega_2 (a_1^+ - a_2^+) \right.$$

$$\left. - i\sqrt{2}\omega_3 (a_1^+ - a_3^+) - ... - i\sqrt{2}\omega_4 (a_1^+ - a_n^+) + \frac{1}{2} a_j^+ a_j^+ \right] | vacuum >, \tag{16}$$

where $\omega = (\omega_2, \omega_3, ..., \omega_n)$. In terms of the eigenvector of the momentum operator $\hat{P}_i$,

$$| p >_i = \pi^{-1/4} \exp\left( -\frac{1}{2} p^2 + i\sqrt{2} p a_i^+ + \frac{1}{2} a_i^{+2} \right) | 0 >_i, \tag{17}$$

we can rewrite Eq.(16) as

$$\frac{1}{(2\pi)^{(n-1)/2}} \int_{-\infty}^{\infty} | p, \chi > \exp[-i(\omega_2\chi_2 + \omega_3\chi_3 + ... + \omega_n\chi_n)] d\chi_2 d\chi_3 ... d\chi_n$$

$$= | p/n - \sum_{k=2}^{n} \omega_k >_1 \otimes | p/n + \omega_2 >_2 \otimes | p/n + \omega_3 >_3 \otimes ... \otimes | p/n + \omega_n >_n. \tag{18}$$

It then follows that

$$| p, \chi > = \frac{1}{(2\pi)^{(n-1)/2}} \int_{-\infty}^{\infty} | p/n - \sum_{k=2}^{n} \omega_k >_1 \otimes | p/n + \omega_2 >_2 \otimes$$

$$... \otimes | p/n + \omega_n >_n \exp\left[ i \sum_{k=2}^{n} \omega_k \chi_k \right] d\omega_2 d\omega_3 ... d\omega_n$$

$$= \frac{1}{(2\pi)^{(n-1)/2}} \exp\left[ -\frac{ip}{n} \sum_{k=2}^{n} \chi_k \right] \int_{-\infty}^{\infty} | p - \sum_{k=2}^{n} \omega_k >_1 \otimes | \omega_2 >_2 \otimes$$

$$... \otimes | \omega_n >_n \exp\left[ i \sum_{k=2}^{n} \omega_k \chi_k \right] d\omega_2 d\omega_3 ... d\omega_n, \tag{19}$$

where $| p - \sum_{k=2}^{n} \omega_k >_1$, $| \omega_2 >_2, \dots, | \omega_n >_n$ are all momentum eigenstates. Eq.(19) is the decomposition of $| p, \chi >$ in the momentum representation. Applying the operator

$\exp(i\hat{X}_1\sum_{k=2}^{n}\hat{P}_k)$ to the state $|p,\chi>$, we have

$$\exp(i\hat{X}_1\sum_{k=2}^{n}\hat{P}_k)|p,\chi>$$

$$=\frac{1}{(2\pi)^{(n-1)/2}}\exp\left[-\frac{ip}{n}\sum_{k=2}^{n}\chi_k\right]\int_{-\infty}^{\infty}\exp\left(i\hat{X}_1\sum_{k=2}^{n}\omega_k\right)|p-\sum_{k=2}^{n}\omega_k>_1\otimes|\omega_2>_2\otimes$$

$$...\otimes|\omega_n>_n\exp\left[i\sum_{k=2}^{n}\omega_k\chi_k\right]d\omega_2 d\omega_3...d\omega_n$$

$$=\frac{1}{(2\pi)^{(n-1)/2}}\exp\left[-\frac{ip}{n}\sum_{k=2}^{n}\chi_k\right]|p>_1\otimes\int_{-\infty}^{\infty}|\omega_2>_2\otimes$$

$$...\otimes|\omega_n>_n\exp\left[i\sum_{k=2}^{n}\omega_k\chi_k\right]d\omega_2 d\omega_3...d\omega_n$$

$$=\exp\left[-\frac{ip}{n}\sum_{k=2}^{n}\chi_k\right]|p>_1\otimes|-\chi_2>_2\otimes|-\chi_3>_3\otimes...\otimes|-\chi_n>_n, \qquad (20)$$

where $|p>_1$ is the eigenstate of momentum operator $\hat{P}_1$, $|-\chi_2>_2,\ldots,|-\chi_n>_n$ are the eigenstates of the coordinate operators $\hat{X}_2$, $\hat{X}_3,\ldots,\hat{X}_2$ respectively. It then follows from Eq.(20) that

$$|p,\chi>=$$

$$\exp\left[-\frac{ip}{n}\sum_{k=2}^{n}\chi_k\right]\exp\left(-i\hat{X}_1\sum_{k=2}^{n}\hat{P}_k\right)|p>_1\otimes|-\chi_2>_2\otimes|-\chi_3>_3\otimes...\otimes|-\chi_n>_n. \qquad (21)$$

We call $\exp(i\hat{X}_1\sum_{k=2}^{n}\hat{P}_k)$ the entangling operator as it entangles the momentum eigenstate $|p>_1$ and the coordinate eigenstates $|-\chi_2>_2$, $|-\chi_3>_3,\ldots,|-\chi_n>_n$ to the n-mode entangled state $|p,\chi>$.

In the same way, we can obtain the decomposition of $|p,\chi>$ in the coordinate representation

$$|p,\chi>=\frac{1}{\sqrt{2\pi}}\exp\left[-\frac{ip}{n}\sum_{k=2}^{n}\chi_k\right]\int_{-\infty}^{\infty}|x>_1\otimes|x-\chi_2>_2\otimes|x-\chi_3>_3\otimes$$

$$...\otimes|x-\chi_n>_n e^{ixp}dx, \qquad (22)$$

where $|x>_1$, $|x-\chi_2>_2,\ldots,|x-\chi_n>_n$ are all coordinate eigenstates.

Similarly, we can also construct the common eigenvector $|\chi, p_2, p_3,..., p_n>$, for simplicity denoted as $|\chi, p>$, of mutually commutative operators $\sum_{k=1}^{n}\hat{X}_k$, and $\hat{P}_1-\hat{P}_i$, $(i=2,3,...,n)$

$$|\chi, p>=\frac{1}{\sqrt{n}\pi^{n/4}}\exp\left[-\frac{1}{4}y'_j\, y'^{*}_j+y'_j\, a_j^{+}-\frac{2}{n}\sum_{k>l=1}^{n}a_l^{+}a_k^{+}+\frac{n-2}{2n}a_j^{+}a_j^{+}\right]|vacuum>, \quad (23)$$

where

$$y'_1=\frac{\sqrt{2}}{n}[\chi+i\sum_{k=2}^{n}p_k], \quad y'_i=y'_1-i\sqrt{2}p_i, \ (i=2,3,...,n). \quad (24)$$

$|\chi, p>$ satisfies the orthogonal relation

$$<\chi, p|\chi'', p''>=\delta(\chi-\chi'')\delta(p_2-p_2'')\delta(p_3-p_3'')...\delta(p_n-p_n''), \quad (25)$$

and the completeness relation

$$\int_{-\infty}^{\infty}d\chi dp_2 dp_3...dp_n\,|\chi, p><\chi, p|=1. \quad (26)$$

Due to

$$[\hat{X}_1-\hat{X}_k,\ \hat{P}_1-\hat{P}_k]=2i, \ (k=2,3,...,n), \quad (27)$$

$$[\sum_{k=1}^{n}\hat{X}_k,\ \sum_{k=1}^{n}\hat{P}_k]=ni, \quad (28)$$

$|p,\chi>$ and $|\chi, p>$ are mutual conjugate states.

### 3. The squeezing transformation of $|p,\chi>$

From the decomposition of $|p,\chi>$ in the coordinate representation, we have

$$\sum_{k=1}^{n}\hat{X}_k\,|p,\chi>$$

$$=\frac{1}{\sqrt{2\pi}}\exp\left[-\frac{ip}{n}\sum_{k=2}^{n}\chi_k\right]\int_{-\infty}^{\infty}\left(n\omega-\sum_{k=2}^{n}\chi_k\right)|\omega>_1\otimes|\omega-\chi_2>_2\otimes$$

$$|\omega-\chi_3>_3\otimes...\otimes|\omega-\chi_n>_n e^{i\omega p}d\omega,$$

$$= -ni\frac{\partial}{\partial p} \mid p,\chi >, \tag{29}$$

which means that in the $\langle p,\chi|$ representation

$$\sum_{k=1}^{n} \hat{X}_k \to ni\frac{\partial}{\partial p} \quad . \tag{30}$$

It then follows from Eqs.(29) and (7) that

$$< p,\chi \mid \frac{1}{n}\hat{P}\sum_{k=1}^{n}\hat{X}_k$$

$$= ip\frac{\partial}{\partial p} < p,\chi \mid = ie^{p'}\frac{\partial p'}{\partial p}\frac{\partial}{\partial p'} < p = e^{p'},\chi \mid = i\frac{\partial}{\partial p'} < p = e^{p'},\chi \mid . \tag{31}$$

From $\exp\left(-\lambda\frac{\partial}{\partial y}\right) f(y) = f(y-\lambda),\ \ y,\lambda \in R$, we have

$$< p,\chi \mid \exp\left[\frac{i\lambda_1}{n}\hat{P}\sum_{k=1}^{n}\hat{X}_k\right]$$

$$= \exp\left(-\lambda_1\frac{\partial}{\partial p'}\right) < p = e^{p'},\chi \mid = \left\langle e^{-\lambda_1}p,\chi\right| . \tag{32}$$

From the decomposition of $\mid p,\chi >$ in the momentum representation, we have

$$\frac{1}{n}(n\hat{P}_2 - \hat{P}) \mid p,\chi > = -i\frac{\partial}{\partial \chi_2} \mid p,\chi > . \tag{33}$$

It then follows from Eqs.(7) and (33) that

$$< p,\chi \mid \frac{1}{n}\hat{Q}_2(n\hat{P}_2 - \hat{P})$$

$$= i\chi_2\frac{\partial}{\partial \chi_2} < p,\chi \mid = ie^{\chi'_2}\frac{\partial \chi'_2}{\partial \chi_2}\frac{\partial}{\partial \chi'_2} < p,\chi_2 = e^{\chi'_2},\chi_3,...,\chi_n \mid$$

$$= i\frac{\partial}{\partial \chi'_2} < p,\chi_2 = e^{\chi'_2},\chi_3,...,\chi_n \mid . \tag{34}$$

Therefore

$$< p,\chi \mid \exp\left[\frac{i\lambda_2}{n}\hat{Q}_2(n\hat{P}_2 - \hat{P})\right]$$

$$= \exp\left(-\lambda_2\frac{\partial}{\partial \chi'_2}\right) < p,\chi_2 = e^{\chi'_2},\chi_3,...,\chi_n \mid = \left\langle p,\chi_2 e^{-\lambda_2},\chi_3,...,\chi_n\right| . \tag{35}$$

Similarly,

$$< p,\chi \,|\, \exp\left[\frac{i\lambda_k}{n}\hat{Q}_k(n\hat{P}_k-\hat{P})\right] = \left\langle p,\chi_2,\chi_3,...,\chi_k e^{-\lambda_k},...,\chi_n\right|,$$

$$k = 3,4,...,n\,. \tag{36}$$

set

$$S_n = \exp\left\{\frac{i}{n}\left[\lambda_1\hat{P}\sum_{k=1}^{n}\hat{X}_k+\sum_{k=2}^{n}\lambda_k\hat{Q}_k(n\hat{P}_k-\hat{P})\right]-\frac{1}{2}\sum_{k=1}^{n}\lambda_k\right\}. \tag{37}$$

Since $\hat{Q}_k\left(n\hat{P}_k-\hat{P}\right)$, ( $k=2,3,...,n$ ) and $\hat{P}\sum_{k=1}^{n}\hat{X}_k$ are mutually commutative, using Eqs.(35)-(37) and (32) we have

$$\left\langle p,\chi_2,\chi_3,...,\chi_n\right| S_n = \exp\left(-\frac{1}{2}\sum_{k=1}^{n}\lambda_k\right)\left\langle e^{-\lambda_1}p,e^{-\lambda_2}\chi_2,e^{-\lambda_3}\chi_3,...,e^{-\lambda_n}\chi_n\right|, \tag{38}$$

which implies that the unitary operator $S_n$ corresponds to the quantum mechanical image of the classical transformation $\mathrm{p}\to e^{\lambda_1}p$ , $\chi_2\to e^{\lambda_i}\chi_i$ , $i=2,...,n$ , in the state $\left|p,\chi_2,\chi_3,...,\chi_n\right\rangle$. Now using the completeness relation (9), we have

$$S_n = \exp\left(-\frac{1}{2}\sum_{k=1}^{n}\lambda_k\right)\int_{-\infty}^{\infty}dpd\chi_2 d\chi_3...d\chi_n \,|\, p,\chi_2,\chi_3,...,\chi_n >< e^{-\lambda_1}p,e^{-\lambda_2}\chi_2,e^{-\lambda_3}\chi_3,...,e^{-\lambda_n}\chi_n \,|$$

$$= \exp\left(\frac{1}{2}\sum_{k=1}^{n}\lambda_k\right)\int_{-\infty}^{\infty}dpd\chi_2 d\chi_3...d\chi_n \,|\, e^{\lambda_1}p,e^{\lambda_2}\chi_2,e^{\lambda_3}\chi_3,...,e^{\lambda_n}\chi_n >< p,\chi_2,\chi_3,...,\chi_n \,|. \tag{39}$$

It then follows that

$$S_n\hat{P}S_n^{-1} = e^{-\lambda_1}\hat{P}\,,\quad S_n\hat{Q}_iS_n^{-1} = e^{-\lambda_i}\hat{Q}_i\,,\ (i=2,3,...,n). \tag{40}$$

We now present the dynamic Hamiltonian for generating such an evolution. For simplicity, we take $\lambda_2=\lambda_3=...=\lambda_n$ . Let the parameter $\lambda_i$ , ( $i=1,2,3,...,n$ ), be time-dependent, we seek the interaction Hamiltonian which can generate the transformation $|\, p,\chi_2,\chi_3,...,\chi_n > \to |\, e^{\lambda_1}p,e^{\lambda_2}\chi_2,e^{\lambda_3}\chi_3,...,e^{\lambda_n}\chi_n >$. For this purpose ,we differentiate (37) with respect to t and obtain

$$i\frac{\partial S_n(t)}{\partial t} = \left[-\frac{i}{2}\frac{\partial\lambda_1}{\partial t}a^+G'\tilde{a}^+ + \frac{i}{2}\frac{\partial\lambda_2}{\partial t}a^+G''\tilde{a}^+ + H.c.\right]S_n(t)\,, \tag{41}$$

where $H.c.$ denotes the Hermitian conjugate,

$$a^{+} = (a_1^{+}, a_2^{+}, ..., a_n^{+}), \tag{42}$$

$$G'_{ij} = 1/n, \quad G''_{ij} = \begin{cases} (n-1)/n, i = j \\ -1/n, i \neq j \end{cases}, \quad (i, j = 1,2,...,n). \tag{43}$$

Recasting Eq.(41) in the standard form for the Schrodinger equation of motion in the interaction picture

$$i\frac{\partial S_n(t)}{\partial t} = H_I(t)\ S_n(t), \tag{44}$$

we obtain the related Hamiltonian

$$H_I(t) = -\frac{i}{2}\frac{\partial \lambda_1}{\partial t} a^{+} G' \tilde{a}^{+} + \frac{i}{2}\frac{\partial \lambda_2}{\partial t} a^{+} G'' \tilde{a}^{+} + H.c., \tag{45}$$

where $a^{+} = (a_1^{+}, a_2^{+}, ..., a_n^{+})$, $\tilde{a}^{+}$ denotes the transposition. Recently, Pfister and his coworkers have studied multipartite continuous-variable entanglement from concurrent nonlinearities[17]. Their Hamiltonian is given by

$$H_I(t) = i\beta\chi \sum_{i=1}^{n} \sum_{j>i}^{n} a_i^{+} a_j^{+} + H.c., \tag{46}$$

where $\chi$ is the nonlinear coupling coefficient and $\beta$ the real, assumed undepleted, coherent pump field amplitude. Solving the Heisenberg equations for the fields gives asymmetric squeezing rates

$$\sum_{i=1}^{n} \hat{P}_i(t) = \exp[-(n-1)\beta\chi t] \sum_{i=1}^{N} \hat{P}_i, \tag{47}$$

$$\hat{X}_1(t) - \hat{X}_j(t) = \exp(-\beta\chi t)(\hat{X}_1 - \hat{X}_j), \quad (j = 2,...,n), \ t \geq 0 \tag{48}$$

It is easy to see that if we set $\partial\lambda_1/\partial t$ to be $(n-1)\partial\lambda_2/\partial t = -(n-1)\beta\chi$ in Eq.(45), we obtain Eq.(46).

Let $\lambda_1 = \lambda_2 = ... = \lambda_n = \lambda$. Substituting $\hat{X}_i = (a_i + a_i^{+})/\sqrt{2}$ and $\hat{P}_i = (a_i - a_i^{+})/i\sqrt{2}$ into Eq.(37) we obtain

$$S'_n = \exp[\lambda(K_{+} - K_{-})] = \exp(K_{+}\tanh\lambda)\exp(2K_0 \ln\sec h\lambda)\exp(-K_{-}\tanh\lambda), \tag{49}$$

where

$$K_{+} = (K_{-})^{+} = \frac{1}{2} a^{+} G \tilde{a}^{+}, \tag{50}$$

$$K_0 = \frac{1}{2} a^{+} \tilde{a} + \frac{1}{4} n, \tag{51}$$

$$G_{ij}=\begin{cases}1-2/n, i=j\\ -2/n, i\neq j\end{cases}, \quad i,j=1,2,...,n, \tag{52}$$

$$a=(a_1,a_2,...,a_n), \qquad a^+=(a_1^+,a_2^+,...,a_n^+), \tag{53}$$

$\tilde{a}$ and $\tilde{a}^+$ denote the transpose of $a$ and $a^+$, respectively. It can be proved that $K_\pm$, $K_0$ satisfy the commutation relation of the generators of the $SU(1,1)$ Lie algebra

$$[K_-,K_+]=2K_0, \quad [K_0,K_+]=K_+, \quad [K_0,K_-]=-K_-. \tag{54}$$

Let

$$\hat{X}=\frac{1}{\sqrt{2n}}\sum_{k=1}^{n}\hat{X}_k, \qquad \hat{Y}=\frac{1}{\sqrt{2n}}\sum_{k=1}^{n}\hat{P}_k, \qquad [\hat{X},\hat{Y}]=\frac{i}{2}, \tag{55}$$

Using the operator identity

$$e^{A}Be^{-A}=\sum_{k=0}^{\infty}\frac{1}{k!}[A,B]_k, \tag{56}$$

where $[A,B]_0=B$, $[A,B]_k=[A,[A,B]_{k-1}]$ and Eq.(37) we have

$$S'^{-1}_n\hat{X}S'_n=e^{-\lambda}\hat{X}, \quad S'^{-1}_n\hat{Y}S'_n=e^{\lambda}\hat{Y}. \tag{57}$$

Applying the operator $S'_n$ to the N-mode vacuum state we obtain the n-mode squeezed vacuum state

$$S'_n\left|vacuum\right\rangle = \mathrm{sec}\,h^{n/2}\lambda\,\exp(K_+\tanh\lambda)|\,vacuum>. \tag{58}$$

The quantum fluctuation of the operator square in the state $S'_n\left|vacuum\right\rangle$ is

$$\left\langle(\Delta\hat{X})^2\right\rangle=\frac{1}{4}e^{-2\lambda} \qquad \left\langle(\Delta\hat{Y})^2\right\rangle=\frac{1}{4}e^{2\lambda}, \tag{59}$$

where $\Delta\hat{X}$ stands for $\hat{X}-<\hat{X}>$, $<\hat{X}>$ is the expectation. Thus the minimum uncertainty relation still remains

$$\Delta X\Delta Y\equiv\sqrt{\left\langle(\Delta X)^2\right\rangle\left\langle(\Delta Y)^2\right\rangle}=\frac{1}{4}, \tag{60}$$

In the limit of infinite squeezing ( in the sense that $\lambda\to\infty$), we have

$$S'_n\left|vacuum\right\rangle|_{\lambda\to\infty}\sim\exp(K_+)|\,vacuum>\sim\int dp\,|\,p,p,...,p>. \tag{61}$$

According to van Loock and Braunstein's method[16], this state can be generated by applying the beam splitter operator $\hat{N}_{1...N}$ to a zero-position eigenstate in mode 1 and n-1 zero-momentum eigenstates in modes 2 through n.

If $\lambda_1 = \lambda, \lambda_2 = \lambda_3 = ... = \lambda_n = 0$, from Eq.(37), we have

$$S''_n = \exp[-\lambda(K'_+ - K'_-)] = \exp(-K'_+ th\lambda)\exp(2K'_0 \ln \sec h\lambda)\exp(K'_- th\lambda), \quad (62)$$

where

$$K'_+ = (K'_-)^+ = \frac{1}{2} a^+ G' \tilde{a}^+, \quad (63)$$

$$K'_0 = \frac{1}{2} a^+ G' \tilde{a} + \frac{1}{4}. \quad (64)$$

They satisfy the commutation relation

$$[K'_-, K'_+] = 2K'_0, \quad [K'_0, K'_+] = K'_+, \quad [K'_0, K'_-] = -K'_-. \quad (65)$$

We call $S''_n$ the n-mode one-side squeezing operator.

If we set $\lambda_1 = 0, \lambda_2 = \lambda_3 = \lambda_4 = \lambda$ in Eq.(37), another N-mode one-side squeezing operator can be obtained,

$$S'''_n = \exp[\lambda(K''_+ - K''_-)] = \exp(K''_+ th\lambda)\exp(2K''_0 \ln \sec h\lambda)\exp(K''_- th\lambda), \quad (66)$$

where

$$K''_+ = (K''_-)^+ = \frac{1}{2} a^+ G'' \tilde{a}^+, \quad (67)$$

$$K''_0 = \frac{1}{2} a_j^+ G'' a_j + \frac{1}{4}(n-1), \quad (68)$$

$K''_\pm$, $K''_0$ also satisfy the commutation relation of the generators of the $SU(1,1)$ Lie algebra

$$[K''_-, K''_+] = 2K''_0, \quad [K''_0, K''_+] = K''_+, \quad [K''_0, K''_-] = -K''_-. \quad (69)$$

4.Summary

In summary, we have generalized the concept of two-particle EPR entangled state to the case of n particle systems and constructed the common eigenvector $|p,\chi>$ of n compatible operators $\hat{Q}_i = \hat{X}_1 - \hat{X}_i$, ($i = 2,3,...,N$), and the total momentum operator $\hat{P} = \sum_{k=1}^{n} \hat{P}_k$. By virtue of the technique of integration within an ordered product (IWOP) of operators we have proved that $|p,\chi>$ is able to make up a multipartite entangled state representation and analysed its entangling properties. By means of the decompositions in coordinate and momentum representation, we have studied the quantum mechanical operator corresponding to a classical transformation $p \to e^{\lambda_1} p$, $\chi_2 \to e^{\lambda_i}\chi_i$, $i = 2,...,n$, in the state $|p,\chi\rangle$ and obtained three n-mode realizations of the SU(1,1) Lie algebra as well as the corresponding n-mode squeezed states. We have also discussed their quantum optical implementation. This state is useful because it is able to make up a quantum mechanical representation. Its further applications in the study of multipartite teleportation, quantum dense coding and entangled fractional Fourier transformation are under consideration.